\documentclass[11pt]{article}
\usepackage{histocrypt}
\usepackage{mathptmx}
\usepackage{url}
\usepackage{latexsym}

\usepackage{graphicx}

\usepackage{hyperref}

\usepackage{tikz} 
\usetikzlibrary{arrows.meta, positioning, fit, backgrounds, calc}

\usepackage{amsmath}
\usepackage{adjustbox}
\usepackage{multirow}

\usepackage{textcomp}
\usepackage{arydshln}

\title{Attention-Augmented LSTMs for Automatic Homophonic Ciphertext Decipherment}

\author{ 
  Micaella Bruton \\
  Stockholm University \\
  {\tt first.last@ling.su.se} \\\And
  Meriem Beloucif \\
  Uppsala University \\
  {\tt first.last@lingfil.uu.se} \\\And
  Beáta Megyesi \\
  Stockholm University \\
 {\tt first.last@ling.su.se}\\
  \
  }

\date{}

\begin{document}

\maketitle
\begin{abstract}

    

    Homophonic substitution ciphers replace each plaintext letter with one of several possible ciphertext codes, deliberately weakening letter-frequency patterns and making automated decipherment difficult. This paper evaluates whether an attention-augmented Long Short-Term Memory (LSTM) model can learn such mappings in a historically motivated shared-key setting: all ciphertexts draw from the same known homophonic code pool, while individual keys use different consistent subsets of that pool. Using synthetic ciphertexts generated with \texttt{ChronoFidelius} from historical English and Swedish texts dated 1500--1899, we test performance across ciphertext lengths, centuries, variable-length codes, and simulated transcription errors. Models are trained only on aligned ciphertext--plaintext pairs, without external language models, frequency statistics, or key-search heuristics. Results show near-perfect character-level decryption accuracy across both languages and all periods, including short and noisy ciphertexts. The model also fails predictably on ciphertexts outside the shared pool, indicating that it functions as a practical tool for decipherment and key-space verification when key reuse is suspected.

\end{abstract}

\section{Introduction}
Homophonic substitution ciphers represent a significant class of historical ciphers that can pose considerable challenges for automated cryptanalysis. By allowing multiple ciphertext symbols or codes to represent a single plaintext character, these systems deliberately obscure frequency distributions, reducing the effectiveness of standard frequency-based analyses. As a result, their decryption has traditionally required more sophisticated techniques, often combining heuristic search with detailed language modelling. Homophonic ciphers appear frequently in early modern diplomatic correspondence and military archives and continue to be encountered in large-scale digitization efforts \cite{megyesi_WhatWasEncoded_2022}. Figure~\ref{fig:example} shows a representative example of a homophonic ciphertext instance used in this study.

\begin{figure*}[h]
\centering
\resizebox{\linewidth}{!}{  
\begin{tabular}{|c|c|c|c|c|c|c|c|c|c|c|c|}
\hline \hline
\textbf{CT} & \texttt{0675} & \texttt{4770} & \texttt{454} & \texttt{0086} & \texttt{443} & \texttt{1987} & \texttt{332} & \texttt{6592} & \texttt{5543} & \texttt{336} & \texttt{9499} \\
\hline
\textbf{PT} & \texttt{T} & \texttt{H} & \texttt{E} & \texttt{L} & \texttt{A} & \texttt{W} & \texttt{E} & \texttt{W} & \texttt{<ERR>} & \texttt{A} & \texttt{S} \\
\hline \hline
\end{tabular}
}
\caption{Aligned excerpt from a homophonic digit cipher example from the English test dataset. Each ciphertext code maps to a single plaintext character, while multiple distinct codes may correspond to the same character across the sequence. The example includes variable-length code and an explicit transcription error marker (\texttt{<ERR>}).}
\label{fig:example}
\end{figure*}

Past automated cryptanalytic approaches have been dominated by search-based methods. Techniques such as nested hill climbing, simulated annealing, beam search, and Bayesian inference iteratively optimize candidate keys using language-dependent fitness functions, typically based on character or word $n$--gram statistics \cite{ravi_BayesianInferenceZodiac_2011,dhavare_EfficientCryptanalysisHomophonic_2013,nuhn_ImprovedDeciphermentHomophonic_2014,kopal_CryptanalysisHomophonicSubstitution_2019,lasry_DecipheringPapalCiphers_2021}. While these methods have achieved notable successes, including the automatic decipherment of historically famous ciphers such as the Zodiac-408, they often rely on language-specific resources. More recently, neural sequence models have been applied to the automatic classical cipher decryption problem. These models reformulate cryptanalysis as a sequence-to-sequence prediction task, learning the decryption function directly from data rather than relying on explicit cryptanalytic heuristics \cite{ahmadzadeh_NovelDynamicAttack_2021,park_AutomatedClassicalCipher_2023}. While these approaches have demonstrated strong performance on text based decryption of simple substitution and polyalphabetic ciphers, their application to homophonic substitution remains relatively underexplored.

This paper investigates whether a neural sequence model based on Long Short-Term Memory (LSTM) networks augmented with attention can serve as a practical tool for the automatic decryption of homophonic substitution ciphers sharing a key space\cite{hochreiter_LongShortTermMemory_1997}. This is relevant as real world ciphertexts often reuse subsets of a larger shared key in real-world historical ciphers, such as the Mary Stuart letters decrypted in 2023 \cite{lasry_DecipheringMaryStuarts_2023}. The main focus is on evaluating:
\begin{itemize}
    \item whether LSTMs with attention can learn a shared homophonic key space reliably enough to serve as a decipherment tool, meeting the standards of robustness, generality, and reliability expected in historical cryptanalysis without the need for additional language-specific tools such as language modelling;
    \item how performance varies diachronically in the presence of language variation over time;
    \item how different noise variants that reflect real-world archival challenges affect model performance.
\end{itemize}

To this end, an experimental study using synthetic homophonic ciphertexts, derived from historical English and Swedish texts spanning the years 1500--1899 was conducted. Models were trained on temporally balanced data and evaluated separately for each century to explicitly measure diachronic generalization.  No additional language models or language-specific tools are used to aid the readability of the output plaintext, ensuring that the models learn decipherment behaviour solely from the ciphertext inputs. Multiple ciphertext variants, including variable-length cipher codes and simulated transcription errors, are introduced to assess robustness under adverse conditions. Finally, to address data scarcity and potential overfitting, 5-fold cross-validation experiments are performed on shorter ciphertexts, which are more abundant and typically more challenging to decrypt. Overall, this study provides an empirically grounded evaluation of LSTMs as assistive tools for historical cryptanalysis, emphasizing robustness, generalizability, and practical applicability within a shared key space setting.

\section{Related Work}
This section provides an overview of both traditional and neural approaches in automated cryptanalysis.

\subsection{Traditional Automated Cryptanalysis Approaches}
Automated homophonic cryptanalysis has traditionally relied on heuristic search techniques guided by language models. One widely used approach is nested hill climbing, which generalizes methods originally developed for simple substitution ciphers by iteratively modifying candidate keys and evaluating them using $n$-gram-based fitness functions \cite{dhavare_EfficientCryptanalysisHomophonic_2013}. While effective, hill climbing is prone to local optima, particularly in the large and irregular search spaces induced by homophonic mappings.

Simulated annealing addresses this limitation by introducing stochastic acceptance of suboptimal moves, allowing the search process to escape local maxima. This approach has been successfully implemented in tools such as CrypTool~2 and has successfully solved real-world ciphers, such as Zodiac-408 and letters written by Mary Stuart, Queen of Scots \cite{kopal_CryptanalysisHomophonicSubstitution_2019,lasry_DecipheringMaryStuarts_2023}. 

Beam search methods provide an alternative strategy by maintaining a fixed number of candidate hypotheses at each stage of decipherment. Nuhn et al. \shortcite{nuhn_ImprovedDeciphermentHomophonic_2014} introduced a beam search approach that optimizes the order in which cipher symbols are deciphered, achieving the first automatic solution of the Beale Cipher (Part~2) and improved results on the Zodiac-408. 

Other methods include evolutionary and probabilistic approaches. Genetic algorithms have been applied to homophonic ciphers by evolving word placements under linguistic constraints, successfully deciphering Zodiac-408 \cite{oranchak_EvolutionaryAlgorithmDecryption_2008}. Bayesian inference methods combine $n$-gram language models with sparse priors and efficient sampling techniques, providing robustness to misspellings and missing word boundaries \cite{ravi_BayesianInferenceZodiac_2011}. While effective, these methods share a common limitation; they rely on rich linguistic resources and language-specific tools, like $n$-gram statistics, for optimal performance.

Hierarchical clustering has also been used in historical cryptanalysis, primarily as an exploratory tool. By clustering ciphertext symbols based on contextual similarity, researchers identified groups of potential homophones in the Copiale cipher, providing a critical starting point for its eventual decipherment \cite{knight_CopialeCipher_2011,lehofer_ApplyingHierarchicalClustering_2022}. This technique is typically used in conjunction with other cryptanalytic methods, rather than as a standalone solution.

\subsection{Neural Approaches \& Workflows}
Recent work has explored neural networks as an alternative method for classical cipher decryption. Ahmadzadeh et al. \shortcite{ahmadzadeh_NovelDynamicAttack_2021} proposed an attention-based LSTM encoder--decoder architecture that treats decipherment as a sequence prediction task. Using modern English corpora to generate ciphertexts, they achieved near-perfect accuracy on shift, Vigenère, and simple substitution ciphers, demonstrating that LSTMs with attention can learn decryption functions without explicit frequency analysis or key information. 

Neural models have also found success for cipher type detection \cite{leierzopf_DetectionClassicalCipher_2021,bastian_EnhancingClassicalCipher_2025}. Unsupervised approaches based on generative adversarial networks (GANs) have also been proposed for decryption. CipherGAN and its extensions learn mappings between plaintext and ciphertext without paired training data, while UC-GAN extends this framework to multiple cipher types within a single model \cite{ahmadzadeh_DeepBidirectionalLSTMGRU_2022,park_AutomatedClassicalCipher_2023}. While all of these models show promising results, their experiments were limited to clean ciphertexts, modern language data, and simple cipher types, such as Caesar shift ciphers, which are less complex than homophonic substitution.

Modern cryptanalytic workflows increasingly integrate decipherment with transcription and segmentation directly on images. Projects such as DECRYPT have combined computer vision models with cryptanalysis pipelines to process handwritten historical documents directly \cite{megyesi_DecryptionHistoricalManuscripts_2020,heder_SupportingHistoricalCryptology_2024}. Recent work has also explored joint transcription--decipherment models to reduce error propagation \cite{Aldarrab:2017,yin_DeciphermentHistoricalManuscript_2019,bermudezgranados_DirectDeciphermentTranscription_2024}. These efforts highlight the growing need for decipherment methods that remain robust under noisy and heterogeneous input conditions.

\subsection{Positioning of the Present Work}
In contrast to prior neural work, the present study focuses explicitly on homophonic substitution ciphers under historically motivated conditions, including diachronic variation, variable-length cipher codes, and simulated transcription errors. This work examines whether a bidirectional LSTM (biLSTM) augmented with multihead attention can meet the robustness and reliability required for the automatic decryption of historical homophonic ciphers. By systematically varying language, century, ciphertext length, and noise conditions---factors often idealized or excluded in prior neural work---this study complements existing cryptanalytic methods and provides empirical evidence for the practical applicability of neural sequence models in historical cryptanalysis.

\section{Method}
\subsection{Problem Formulation}
The decryption of homophonic substitution ciphers is formulated as a supervised sequence-to-sequence prediction problem where each plaintext character may correspond to multiple ciphertext codes. Given a ciphertext sequence
\[
C = (x_1, x_2, \dots, x_T),
\]
the objective is to predict the corresponding plaintext sequence:
\[
P = (y_1, y_2, \dots, y_T).
\]
The task is evaluated at the character level, excluding padding symbols introduced for batching. Crucially, all ciphertexts in this study are generated from a shared homophonic key space, meaning that each ciphertext code maps deterministically to the same plaintext character across the entire dataset. Individual keys differ only in which subset of the shared pool is assigned to each plaintext character, but no code is ever assigned to more than one plaintext character. For example, if the shared key pool for \textsc{'a' = \{1, 2, 3, 4\}}, then ciphertext\_a might use \textsc{'a' = \{1, 2, 3\}}, while ciphertext\_b uses \textsc{'a' = \{2, 3, 4\}}. Though independently these keys appear different, both were encrypted using the same shared key space and only differ in the homophones selected for each plaintext representation.

This distinguishes the task from general homophonic cryptanalysis, where keys are independently assigned and codes may represent different plaintext characters across documents. For example, in ciphertext\_a \textsc{'a' = \{1, 2, 3\}}, but ciphertext\_b uses \textsc{'b' = \{1, 2, 3\}}. Rather than performing independent key recovery per ciphertext, the model learns a stable dataset-wide many-to-one mapping, a structurally different and more tractable problem whose difficulty is governed by the consistency of the shared space rather than per-instance key complexity.

Unlike traditional cryptanalytic methods, no explicit key search, frequency analysis, or language-specific functions are used. Instead, the model learns decryption mappings during training directly from aligned ciphertext–plaintext pairs.

\subsection{Dataset and Experimental Design}
\begin{table}[t]
    \centering
    \begin{tabular}{c|r|r|r} \hline \hline
        Language & Training  & Validation & Testing\\ \hline
        English  & 965,344 & 120,656 & 120,692 \\ \hline
        Swedish  & 272,860 & 34,100  & 34,128 \\ \hline \hline
    \end{tabular}
    \caption{Dataset summary. Plaintext sequence counts are shown; ciphertext sequence counts are scaled by four. An equal number of examples for each year range are included within these totals.}
    \label{tab:dataset-summary}
\end{table} 

All experiments use a synthetic dataset of homophonic substitution ciphers distributed via HuggingFace\footnote{\href{https://huggingface.co/collections/mbruton/homophonic-ciphertexts}{huggingface.co/collections/mbruton/homophonic-ciphertexts}}. The English and Swedish language portions are used; data spans the years 1500--1899 and is divided into consecutive 100-year intervals. 

Plaintext data was sourced from the \textit{HistCorp} corpus, spanning multiple genres including religious, legal, and literary \cite{pettersson_HistCorpCollectionHistorical_2018}. Plaintext sequences have had all spacing and punctuation removed, reflecting standard historical encryption practices.

Ciphertext data was generated using \texttt{ChronoFidelius}\footnote{\href{https://github.com/mbruton0426/ChronoFidelius}{github.com/mbruton0426/ChronoFidelius}} \cite{bruton_StatisticsNeuralNetworks_2025}. All ciphers are encrypted using monoalphabetic homophonic substitution where codes were randomly assigned within a fixed, shared homophonic space. Each individual key is unique, while representing a subset of a larger homophonic key space. 
This design is intentional as \texttt{ChronoFidelius} can be used to generate synthetic training data from any known key space, and the resulting models can then be applied to unknown ciphertexts to determine whether they share that space. Examples employing variable-length code used 3-digit numbers for vowels and 4-digit numbers for consonants, while in those without, 4-digit numbers were used for all codes. Train, validation, and test splits contain entirely disjoint plaintexts, ensuring that no plaintext seen during training appears in evaluation. Four different ciphertexts are produced for each plaintext: 
\begin{itemize}
    \item clean ciphertext;
    \item ciphertext with 5\% random character-level substitutions or insertions simulating transcription noise;
    \item ciphertext with variable-length code simulating increased complexity;
    \item ciphertext combining both noise types.
\end{itemize}

Both languages include six plaintext sequence lengths; 50, 200, 400, 600, 800, and 1,000 characters. Ciphertexts were extended by up to 2, 10, 20, 30, 40, and 50 codes respectively if the variant included transcription noise to accommodate inserted characters.

Training data is sampled using a temporally balanced strategy, with equal representation from each century. Separate models are trained for each language--length combination. Standard train/validation/test splits provided by the dataset were used; summary of available data is available in Table \ref{tab:dataset-summary}.

\subsection{Model Architecture}
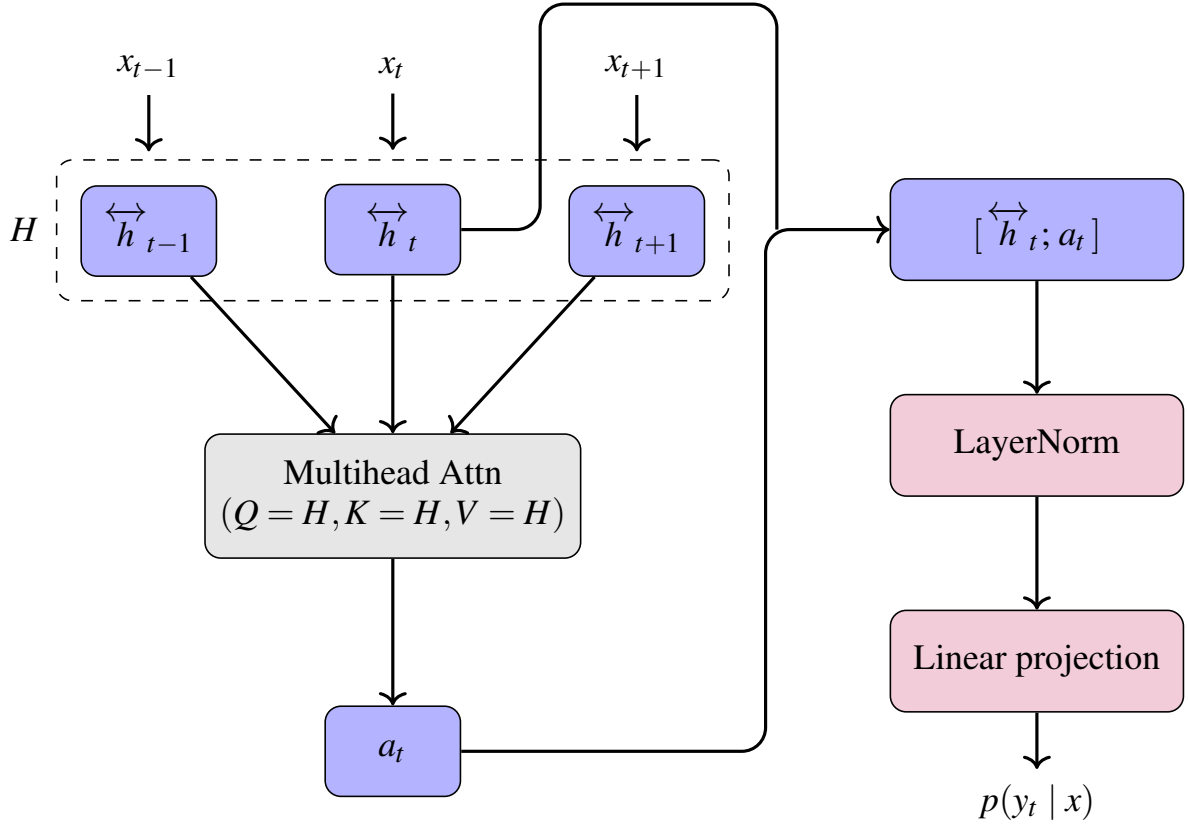
\begin{figure*}[t]
\centering
\resizebox{\linewidth}{!}{  
\begin{tikzpicture}[
    font=\footnotesize,
    state/.style={draw, rounded corners, minimum width=1.2cm, minimum height=0.8cm, fill=blue!30},
    attn/.style={draw, rounded corners, fill=gray!20, minimum width=3.0cm, minimum height=1.1cm},
    block/.style={draw, rounded corners, minimum width=2.6cm, minimum height=0.9cm, fill=purple!20},
    stateblock/.style={draw, rounded corners, minimum width=2.6cm, minimum height=0.9cm, fill=blue!30},
    arrow/.style={->, thick},
    every node/.style={align=center}
]

\node (x1) {$x_{t-1}$};
\node[right=1.5cm of x1] (x2) {$x_t$};
\node[right=1.5cm of x2] (x3) {$x_{t+1}$};

\node[state, below=0.8cm of x1] (h1) {$\overleftrightarrow{h}_{t-1}$};
\node[state, below=0.8cm of x2] (h2) {$\overleftrightarrow{h}_t$};
\node[state, below=0.8cm of x3] (h3) {$\overleftrightarrow{h}_{t+1}$};

\draw[arrow] (x1) -- ($(h1.north)+(0,0.3cm)$);
\draw[arrow] (x2) -- ($(h2.north)+(0,0.3cm)$);
\draw[arrow] (x3) -- ($(h3.north)+(0,0.3cm)$);

\node[left=2.4cm of h2] {$H$};

\node[attn, below=1.4cm of h2] (attnblock) {Multihead Attn\\
$(Q=H,K=H,V=H)$};

\foreach \i in {h1,h2,h3}
  \draw[arrow] (\i) -- (attnblock);

\node[state, below=1.3cm of attnblock] (a2) {$a_t$};
\draw[arrow] (attnblock) -- (a2);

\node[stateblock, right=3.8cm of h2] (concat) {\\$[\,\overleftrightarrow{h}_t;\, a_t\,]$};

\coordinate (merge) at ($(concat.west)+(-1.0,0)$);

\coordinate (wp1) at ($(h2.east)+(0.7,0)$);     
\coordinate (wp2) at ($(wp1)+(0,2.0)$);         
\draw[thick, rounded corners=6pt]
  (h2.east) -- (wp1) -- (wp2) -| (merge);
\draw[arrow, rounded corners=6pt] (a2.east) -- ++(2.7,0) |- (concat.west);

\node[block, below=1.0cm of concat] (norm) {LayerNorm};
\node[block, below=1.0cm of norm] (linear) {Linear projection};

\draw[arrow] (concat) -- (norm);
\draw[arrow] (norm) -- (linear);

\node[below=0.5cm of linear] (out) {$p(y_t \mid x)$};
\draw[arrow] (linear) -- (out);

\begin{pgfonlayer}{background}
\node[draw, dashed, rounded corners, fit=(h1)(h2)(h3), inner sep=6pt] (Hbox) {};
\end{pgfonlayer}
\end{tikzpicture}
}
\caption{Architecture of the homophonic decryption model.}

\label{fig:lstm_multihead}
\end{figure*} 

The decryption model consists of a biLSTM network augmented with a multihead self-attention mechanism. Each ciphertext symbol $x_t$ is first mapped to a learned embedding vector and processed by a multi-layer biLSTM, producing context-aware hidden states $\overleftrightarrow{h}_t$ that capture both left- and right-context dependencies. Dropout is applied to the LSTM outputs to mitigate overfitting.

To address the homophonic nature of the cipher and the presence of long-range dependencies, a multihead self-attention mechanism is applied over the full sequence of hidden states $H$. For each position $t$, the attention mechanism produces an attention vector $a_t$ that dynamically reweigh relevant positions in the ciphertext sequence. The attention output $a_t$ is concatenated with the corresponding LSTM hidden state and the combined representation is normalized before being passed to a linear projection layer. The final output at each position is a probability distribution over the plaintext character vocabulary, yielding $p(y_t \mid x)$. A visualization of the model architecture is shown in Figure~\ref{fig:lstm_multihead}.

\subsection{Training \& Evaluation}
Hyperparameters were fine-tuned using Optuna\footnote{\href{https://github.com/optuna/optuna}{github.com/optuna/optuna}} (50 trials × 10 epochs per model). 
"Err" indicates training with injected transcription errors, while "Mix" denotes variable-length code inclusion.

Separate models are trained per language and ciphertext variant. All models are trained using cross-entropy loss at the character level. Padding symbols are excluded from the loss computation and all evaluation metrics. Optimization is performed using the AdamW optimizer with weight decay. Gradient clipping is applied to prevent exploding gradients. Mixed-precision training is used where supported to improve computational efficiency. A learning-rate scheduler reduces the learning rate when validation loss plateaus, further stabilizing training. Training proceeds until convergence.

Performance is evaluated using character-level F1 score. Evaluation is performed separately on each century to assess diachronic generalization. Metrics are reported separately for each century, language, ciphertext variant, and text length. 

\paragraph{5-Fold Cross Validation}
To further assess model robustness on short ciphertexts, we conducted 5-fold cross-validation experiments on 50-character ciphertexts. For each language, the dataset was partitioned into five equally sized folds, with four folds used for training and the remaining fold held out for evaluation. This process was repeated such that each fold served as the test set exactly once.

We evaluated four experimental configurations corresponding to different combinations of \texttt{errorUsage} and \texttt{variableCodeUsage} being enabled or disabled during training and inference: T/T, T/F, F/F, and F/T, respectively. Model performance was measured using the F1 score between the ground-truth and predicted plaintext characters. Results are reported as the mean F1 across folds, along with observed variability.

\section{Results}
\begin{table}[b]
\caption{Character-level F1 scores for English and Swedish across ciphertext lengths and noise conditions; averaged over year ranges. `error/mix' indicates transcription errors / variable-length codes (F = absent, T = present).}
    \centering
    \resizebox{\linewidth}{!}{  
    \begin{tabular}{cccccccc}
    \hline \hline
     & text length & 50 & 200 & 400 & 600 & 800 & 1000 \\ \hdashline
    language & error/mix &  &  &  &  &  \\
    \hline
    \multirow[t]{4}{*}{English} & F/F & 1 & 1 & 1 & 1 & 1 & 1 \\
     & F/T & 1 & 1 & 1 & 1 & 1 & 1 \\
     & T/F & 1 & 0.996 & 0.998 & 1 & 1 & 0.993 \\
     & T/T & 1 & 1 & 1 & 1 & 0.996 & 0.996 \\
    \cline{1-8}
    \multirow[t]{4}{*}{Swedish} & F/F & 1 & 1 & 1 & 1 & 1 & 1 \\
     & F/T & 1     & 1 & 1 & 1 & 1 & 1 \\
     & T/F & 0.996 & 0.989 & 0.989 & 0.989 & 0.989 & 0.990 \\
     & T/T & 0.997 & 0.986 & 1 & 1 & 0.990 & 0.990 \\
    \cline{1-7}
    \hline \hline
    \end{tabular}
    }
    \label{fig:results}
\end{table} 

\begin{table}[t]
\centering
\resizebox{\linewidth}{!}{  
\begin{tabular}{ccccccc}
\hline \hline
 &  & year range & 1500-1599 & 1600-1699 & 1700-1799 & 1800-1899 \\ \hdashline
language & text length & error/mix &  &  &  &  \\
\hline
\multirow{24}{*}{\rotatebox{90}{English}} & \multirow[t]{4}{*}{50} & F/F & 1 & 1 & 1 & 1 \\
 &  & F/T & 1 & 1 & 1 & 1 \\
 &  & T/F & .999 & 1 & 1 & 1 \\
 &  & T/T & .999 & 1 & 1 & 1 \\
\cline{2-7}
 & \multirow[t]{4}{*}{200} & F/F & 1 & 1 & 1 & 1 \\
 &  & F/T & 1 & 1 & 1 & 1 \\
 &  & T/F & 0.995 & 0.995 & 0.999 & 0.994 \\
 &  & T/T & 1 & 1 & 1 & 1 \\
\cline{2-7}
 & \multirow[t]{4}{*}{400} & F/F & 1 & 1 & 1 & 1 \\
 &  & F/T & 1 & 1 & 1 & 1 \\
 &  & T/F & 0.998 & 0.998 & 0.997 & 0.998 \\
 &  & T/T & 1 & 1 & 1 & 1 \\
\cline{2-7}
 & \multirow[t]{4}{*}{600} & F/F & 1 & 1 & 1 & 1 \\
 &  & F/T & 1 & 1 & 1 & 1 \\
 &  & T/F & 1 & 1 & 1 & 1 \\
 &  & T/T & 1 & 0.999 & 0.999 & 1 \\
\cline{2-7}
 & \multirow[t]{3}{*}{800} & F/F & 1 & 1 & 1 & 1 \\
 &  & F/T & 1 & 1 & 1 & 1 \\
 &  & T/F & 1 & 0.999 & 1 & 1 \\
 &  & T/T & 0.996 & 0.995 & 0.997 & 0.998 \\
\cline{2-7}
 & \multirow[t]{4}{*}{1000} & F/F & 1 & 1 & 1 & 1 \\
 &  & F/T & 1 & 1 & 1 & 1 \\
 &  & T/F & 0.991 & 0.990 & 0.996 & 0.996 \\
 &  & T/T & 0.995 & 0.995 & 0.997 & 0.997 \\
\hline \hline \hline
\multirow{24}{*}{\rotatebox{90}{Swedish}} & \multirow[t]{4}{*}{50} & F/F & 1 & 1 & 1 & 1 \\
 &  & F/T & 1 & 1 & 1 & 1 \\
 &  & T/F & .992 & .997 & .997 & .999 \\
 &  & T/T & .999 & 1 & .991 & .999 \\
\cline{2-7}
 & \multirow[t]{2}{*}{200} & F/F & 1 & 1 & 1 & 1 \\
 &  & F/T & 1 & 1 & 1 & 1 \\
 &  & T/F & .996  & .998  & 1  & .999  \\
 &  & T/T & 0.982 & 0.992 & 0.985 & 0.983 \\
\cline{2-7}
 & \multirow[t]{3}{*}{400} & F/F & 1 & 1 & 1 & 1 \\
 &  & F/T & 1 & 1 & 1 & 1 \\
 &  & T/F & .992  & 1 & .999 &  .993 \\
 &  & T/T & 1 & 0.999 & 1 & 1 \\
\cline{2-7}
 & \multirow[t]{4}{*}{600} & F/F & 1 & 1 & 1 & 1 \\
 &  & F/T & 1 & 1 & 1 & 1 \\
 &  & T/F & 0.992 & 0.992 & 0.987 & 0.985 \\
 &  & T/T & 1 & 1 & 1 & 1 \\
\cline{2-7}
 & \multirow[t]{4}{*}{800} & F/F & 1 & 1 & 1 & 1 \\
 &  & F/T & 1 & 1 & 1 & 1 \\
 &  & T/F & 0.992 & 0.992 & 0.987 & 0.986 \\
 &  & T/T & 0.993 & 0.993 & 0.987 & 0.987 \\
\cline{2-7}
 & \multirow[t]{4}{*}{1000} & F/F & 1 & 1 & 1 & 1 \\
 &  & F/T & 1 & 1 & 1 & 1 \\
 &  & T/F & 0.994 & 0.993 & 0.988 & 0.987 \\
 &  & T/T & 0.993 & 0.992 & 0.988 & 0.987 \\
\cline{1-7} \cline{2-7}
\hline \hline
\end{tabular}
}
\caption{Character-level F1 scores for English and Swedish across text lengths, year ranges, and noise conditions. `error/mix' denotes transcription errors / variable-length codes (F = absent, T = present).}
\label{tab:compact_year_results}
\end{table} 

Table~\ref{fig:results} summarizes character-level F1 scores for English and Swedish across ciphertext lengths and noise conditions, averaged over all year ranges. Across all configurations, performance is consistently high. For clean ciphertexts (F/F) and ciphertexts containing only variable-length cipher codes without transcription errors (F/T), all models achieve near-perfect performance (F1 $\approx$ 1) for all text lengths in both languages.

Strong performance is also observed throughout for short ciphertexts of length 50. Averaged F1 scores for 50-character inputs remain at or near ceiling across all configurations, indicating that the proposed approach remains effective even in low-context settings. This result is further supported by the 5-fold cross-validation results reported in Table~\ref{tab:cv50}, which demonstrate stable generalization on short sequences across data splits.

When transcription errors are introduced without variable-length codes (T/F), performance remains extremely high, with F1 scores above 0.99 across all lengths. The combination of both transcription errors and variable-length codes (T/T) results in a small decrease in performance, particularly for longer sequences, but F1 remains above 0.99 in nearly all cases. Overall, the results indicate that the proposed models reliably learn homophonic decryption mappings and are highly robust to both structural variation and transcription noise.

Variable-length cipher codes alone (F/T) do not negatively impact performance, indicating that the attention mechanism successfully resolves symbol-level ambiguity introduced by homophonic variation. Transcription errors (T/F) introduce minor degradation, but the effect remains small and consistent across lengths.

Ciphertext length in general does not appear to have an impact on decryption accuracy. This suggests that the model’s ability to leverage contextual information scales well with sequence length, and that performance degradation due to length is minimal under the tested conditions.

The most challenging condition is the combination of transcription errors and variable-length codes (T/T). Nevertheless, even under this setting, decryption accuracy remains high (0.9861-1 F1), demonstrating strong robustness to realistic forms of noise encountered in historical material.

Performance patterns are highly consistent across English and Swedish. No systematic differences between languages are observed, and both exhibit similar sensitivity to noise and sequence length. The detailed per-century results in Table~\ref{tab:compact_year_results} further show that performance remains stable across all evaluated historical periods from 1500 to 1899. No degradation is observed for earlier centuries, indicating that the models generalize across diachronic language variation when trained on temporally balanced data.

The consistently high performance across languages, centuries, and noise conditions suggests that the results are not driven by favourable data partitioning. Performance stability across year ranges and text lengths supports the reliability of the learned decryption behaviour and indicates that the models generalize beyond specific temporal or linguistic subsets of the data.

Inspection of model outputs indicates that most observed errors arise from misidentifying the presence or position of transcription errors rather than from incorrect substitution of one plaintext character for another. In cases where performance falls below perfection, decrypted sequences typically preserve the correct homophonic mappings, with errors corresponding to incorrectly inferred error tokens. This suggests that the primary challenge under noisy conditions lies in error detection rather than in learning or applying the underlying substitution structure.

\paragraph{5-Fold Cross Validation}
Table~\ref{tab:cv50} summarizes the 5-fold cross-validation results on 50-character ciphertexts. Across both English and Swedish datasets, all configurations achieved near-perfect performance, with mean accuracies of at least 0.999 in every case. Several configurations attained near-perfect accuracy across all folds, exhibiting zero variance.

Minor variability was observed in a small number of folds on noisy texts, however, overall performance remained effectively at ceiling. These findings indicate that the proposed approach generalizes reliably across data splits and that strong decryption performance on short ciphertexts is not sensitive to fold selection or training--test partitioning.

\begin{table}[t]
\centering
\resizebox{\linewidth}{!}{  
\begin{tabular}{ccccc}
\hline \hline
 & \multicolumn{4}{c}{Error/Variable Code Inclusion} \\
Language & T/T & T/F & F/F & F/T \\
\hline
English  & $0.999 \pm 0.000$ & $0.999 \pm 0.001$ & $1.000 \pm 0.000$ & $0.999 \pm 0.000$ \\
\hline
Swedish & $0.999 \pm 0.000$ & $0.999 \pm 0.001$ & $1.000 \pm 0.000$ & $0.999 \pm 0.001$ \\
\hline \hline
\end{tabular}
}

\caption{5-fold cross-validation F1 scores on 50-character ciphertexts. Results are reported as mean $\pm$ standard deviation across folds. Columns indicate transcription error / variable-length code inclusion (T = present, F = absent).}
\label{tab:cv50}
\end{table}

\section{Discussion}
The results demonstrate that LSTM sequence models augmented with multihead attention can reliably learn monoalphabetic homophonic decryption mappings under a wide range of historically motivated conditions. Taken together, the findings directly address the three evaluation goals outlined in the Introduction: robustness in the absence of language-specific tools, generalization across diachronic language variation, and resilience to realistic forms of noise encountered in archival material.

High performance across all test cases suggests that the model effectively resolves homophonic ambiguity without relying on explicit frequency analysis or external language models. This distinguishes the approach from traditional heuristic and search-based cryptanalytic methods, which typically depend on language-specific $n$-gram statistics and carefully tuned fitness functions where performance can degrade under noisy or structurally irregular input. These results are expected and theoretically grounded given the shared key space; because each set of ciphertext codes map deterministically to the same plaintext character across the dataset, the model is not performing independent cryptanalysis of each ciphertext. Instead, it learns a stable many-to-one mapping of the underlying fixed key space with tractability scaling with the consistency of the shared space rather than per-instance complexity. Near-perfect performance, including on short ciphertexts, is therefore consistent with theoretical expectations in this setting.

To further validate that models are learning the shared key space rather than superficial patterns, ciphertexts encrypted using keys drawn from outside the shared homophonic pool were evaluated. Under these conditions, model performance collapsed, with F1 scores not exceeding 0.08 across all tested configurations. This predictable failure is itself a meaningful and intended result as it confirms that the model has internalized the shared key space and does not generalize beyond it. In practical terms, this behaviour enables the model to function as a key-space verification tool. When applied to an unknown ciphertext, a near-perfect decryption indicates that the ciphertext shares the known key space; a failed decryption indicates that it does not, providing a useful signal for cryptanalysts even in the absence of a perfect decipherment.

Although performance reductions are observed, mainly under combined noise conditions, these reductions remain limited. In practical settings, the residual errors produced by the model are typically sparse, making them readily identifiable during any subsequent manual analysis. Moreover, many historical ciphers consist of short dispatches, fragments, or partial documents rather than long continuous texts, making performance on shorter ciphertexts of greater practical relevance for historical cryptanalysis than absolute performance on the longest sequences. For such ciphertexts, the model maintains consistently high performance across all tested conditions. This includes very short ciphertexts of 50 characters, for which both standard results and 5-fold cross-validation experiments demonstrate near-perfect and highly stable performance across all noise configurations.

Transcription errors introduce the most substantial challenge. Nevertheless, models maintained strong decryption performance even under these conditions. Inspection of model predictions shows that many errors are primarily attributable to the misidentification of transcription errors, rather than incorrect substitution of plaintext characters. In such cases, the underlying homophonic mapping is preserved, with errors arising from uncertainty about error placement or presence rather than symbol-level confusion. This behaviour is significant given that historical ciphertexts are frequently derived from handwritten sources with imperfect transcriptions.

In the remaining error cases, failures are not randomly distributed across symbols. Instead, they arise from a consistent misidentification of the homophonic mapping for a specific cipher code, which is systematically linked to an incorrect plaintext character. All occurrences of the affected code are decoded as the same incorrect character, rather than being inconsistently or arbitrarily assigned. The excerpt shown in Figure \ref{tab:prediction} shows such a case, where one of homophonic ciphertext codes corresponding to \textsc{C} has been misidentified and consistently decoded as \textsc{I}. This behaviour, observed across sequence lengths and noise conditions, indicates that the model is able to identify a stable, albeit partially incorrect, decryption mapping rather than defaulting to random substitution or lexical guessing under uncertainty. Overall, these results further support the interpretation that the model is genuinely learning the structure of the shared key space.

\begin{table}
    \resizebox{\linewidth}{!}{  
    \centering
    \begin{tabular}{c|c|c|c|c|c|c|c|c|c|c|c|c} \hline \hline
        \textbf{Target}     & T & U & R & N & E & D & T & O & L & Y & C & E \\ \hline
        \textbf{Prediction} & T & U & R & N & E & D & T & O & L & Y & I & E \\ \hline \hline
    \end{tabular}
    }
    \caption{Excerpt from the English evaluation dataset illustrating a stable misidentification of a homophonic cipher code.}    
    \label{tab:prediction}
\end{table}

\paragraph{Why Sequence Models Are Effective for Historical Homophonic Ciphers}
Homophonic ciphers are governed by fixed substitution mappings within individual documents, even though multiple ciphertext symbols may correspond to the same plaintext character. Sequence-based models are well suited to this structure because they can exploit long-range dependencies and contextual regularities across the ciphertext. Once sufficient contextual evidence is available, the underlying substitution mapping becomes learnable without explicit key search.

The inclusion of attention mechanisms further supports this process by allowing the model to dynamically integrate information across the ciphertext sequence, enabling disambiguation when local context alone is insufficient. Importantly, the models operate without access to explicit language models, character frequency statistics, or temporal metadata. This indicates that decryption behaviour is learned directly from the structure of the ciphertext--plaintext relationship within the key space.

The absence of systematic performance degradation across centuries demonstrates that the model does not implicitly rely on language features specific to any particular historical period. Instead, it learns decryption behaviour that generalizes across historical language variation when training data are temporally balanced. This finding reinforces the importance of diachronic coverage in training data and supports the applicability of neural sequence models to historically diverse corpora.

Overall, these results confirm that neural sequence models warrant further exploration as a complementary tool alongside established cryptanalytic techniques. In contrast to both traditional heuristic search methods and prior neural approaches evaluated primarily on clean, modern data, the present study demonstrates that sequence models can operate robustly under historically realistic conditions within a shared key space, with performance stability further confirmed through cross-validation on short ciphertexts, making them particularly suitable for large-scale or heterogeneous archival collections.

\section{Conclusion}
This paper investigated the feasibility of using self-attention-augmented LSTMs for the automatic decipherment of monoalphabetic homophonic ciphers within a shared key space under historically motivated conditions. Through extensive experiments on synthetic ciphertexts spanning multiple languages, centuries, text lengths, and noise scenarios, we demonstrated that such models can reliably learn homophonic substitution mappings without reliance on explicit language models or handcrafted cryptanalytic heuristics.

The results show near-perfect decryption performance for clean ciphertexts and those containing variable-length cipher codes, as well as strong robustness to transcription errors. Even under combined noise conditions, accuracy remains high across all evaluated text lengths and historical periods. In cases where errors persist, they predominantly reflect uncertainty in error identification rather than incorrect recovery of the underlying substitution mapping. Performance stability on short ciphertexts is further supported by cross-validation experiments, indicating that results are not sensitive to particular data partitions. Evaluation on ciphertexts outside the shared homophonic pool resulted in predictable failure, confirming the model genuinely learns the underlying shared key space rather than superficial patterns. These findings enable the use of this method as a key-space verification and decipherment tool for historical cryptanalytic researchers. Researchers who have access to a known key can utilize this method to create synthetic data and train a decipherment model, which can then proactively scan transcripted ciphertexts to determine whether or not they shared the known key space.


Future work will focus on extending this approach toward independent key space decryption, multilingual and language-agnostic decryption, and integrating neural decryption models into end-to-end pipelines that support transcription, decipherment, and key extraction. Within this broader context, the present study establishes a solid foundation for the use of neural sequence models as practical assistive tools for historical cryptanalysis rather than replacements for established expertise.

\section*{Acknowledgments}


\bibliographystyle{histocrypt}
\bibliography{histocrypt}

\end{document}